\documentclass[10pt,twocolumn]{article} 
\usepackage{ol2}
\usepackage{hyperref}
\usepackage{amsmath}
\usepackage{amsfonts}
\usepackage{amsmath}
\usepackage{amssymb}
\begin{document}
\newcommand{\ud}{\mathrm{d}}
\newcommand{\ue}{\mathrm{e}}
\newcommand{\disp}{\displaystyle}
\newcommand{\der}{\partial}
\newcommand{\intl}{\int_{-\infty}^{+\infty}}
\newcommand{\imag}{\mathrm{Im}}
\twocolumn[
\title{\bf Dynamical oscillations in nonlinear optical media}
\author{Theodoros P. Horikis$^1$ and Hector E. Nistazakis$^2$}
\address{$^1$Department of Mathematics, University of Ioannina, Ioannina 45110, Greece\\
$^2$Department of Physics, University of Athens, Panepistimiopolis, Zografos, Athens 15784, Greece }
\begin{abstract}
The spatial dynamics of pulses in Kerr media with parabolic index profile are
examined. It is found that when diffraction and graded-index have opposite signs
propagating pulses exhibit an oscillatory pattern, similar to a breathing behavior.
Furthermore, if the pulse and the index profile are not aligned the pulse
oscillates around the index origin with frequency that depends on the values of the
diffraction and index of refraction. These oscillations are not observed when
diffraction and graded-index share the same sign.
\end{abstract}
]   
\noindent In recent years, great effort has been made in order to explain the
behavior of light beams propagating through interfaces separating optical media
with different nonlinear refractive indices. This interest has been motivated by
different factors that include the analysis of pulse propagation and self-phase
modulation effects in fibers, the close connection with the problem of radiation
mode propagation in the three-dimensional nonlinear Schr\"{o}dinger equation, and
the persistency in the nonlinear wave propagation regime of some properties found
in the corresponding linear problem, such as the occurrence of a parametric
instability under modulation of parameters \cite{longhi,michinel}.

Nonlinear pulse propagation in graded-index optical waveguides is a much studied
problem, both numerically and analytically. In early works, the radial dynamics of
pulses in nonlinear fibers were studied by using the paraxial ray approximation.
However, this was shown to give inaccurate results for the self-phase modulation of
a pulse that propagates in a bulk nonlinear medium (see \cite{karlsson} and
references therein) and a variational method was adopted. Using this approach the
authors in Ref. \cite{karlsson} studied the resulting dynamics of pulses from the
combined effects of spatial diffraction, nonlinearity and parabolic graded index in
radially symmetric fibers. More recently, in Ref. \cite{longhi2} the space-time
dynamics in nonlinear multimode parabolic index optical fibers were studied in the
context of a temporal modulation instability induced by spatial transverse effects.
We extend this approach to include the dynamics of the pulse's center in planar
waveguides. In fact, we find that the motion of the center of the pulse does not
depend on any other parameter of the pulse and it is only determined by the values
of the diffraction and the refractive index.

The nonlinear Schr\"{o}dinger equation appropriately modified to model beam
propagation in graded-index, nonlinear waveguide amplifiers with refractive index
$n(z,x)=n_0+n_1 x^2+n_2 |\psi|^2$ is
\begin{gather}
i\psi_z+\frac{d_0}{2}\psi_{xx}+\frac{1}{2}n_1 x^2\psi+n_2|\psi|^2\psi=0 \label{nls}
\end{gather}
The parameters, $d_0$, $n_1$ and $n_2$ are constant, diffraction ($d_0$) can be
positive or negative and the medium can be anti-guiding ($n_1>0$) or guiding
($n_1<0$). As in Ref. \cite{agrawal2}, both cases are consider in order to examine
the interplay between self-defocusing of light and guiding or anti-guiding and
diffractive effects of the medium.

Interestingly, Eq. (\ref{nls}) is similar to the Gross–-Pitaevskii equation that
describes the dynamics of confined atomic Bose–-Einstein condensates (BECs)
\cite{bec2,bec3}. The crucial differences are that, in BECs, $d_0$ is always
positive and $n_1=-\Omega^2<0$, where $\Omega$ is the normalized harmonic trap
strength. As also shown below, this difference in signs is what gives rise to the
so-called {\em collective oscillations} in the BECs context \cite{abdull1}.
Furthermore, a dissipative variant of this equation has also been used to describe
the behavior of solitons and self-similar waves in nonlinear systems exhibiting
both spatial inhomogeneity and gain or loss at the same time \cite{agrawal,wu}.

Equation (1) can be restated in variational form using the Langrangian density
\[
\mathcal{L}=\frac{i}{2}(\psi_z\psi^*-\psi_z^*\psi)-\frac{d_0}{2}|\psi_x|^2 +
\frac{n_1}{2}x^2|\psi|^2 + \frac{n_2}{2}|\psi|^4
\]
If the injected field is a gaussian beam, in the weakly or moderate nonlinear
regime ($n_2\approx0$) the field remains approximately gaussian and the periodic
variation of the beam parameters along the propagation distance can be calculated
by means of a variational approach \cite{karlsson,longhi2}. Thus, we take
\begin{gather}
\psi(z,x)=A(z)\ue^{-b(z)[x-x_0(z)]^2}\ue^{i\phi(z,x)} \label{profile}
\end{gather}
where $\phi(z,x)=\alpha(z)x^2+\beta(z)x+\gamma(z)$, describes the pulse's phase.
For a given value of $z$, Eq. (\ref{profile}) defines a gaussian beam invariant
along the $y$ direction (planar waveguide geometry) displaced by an amount $x_0$
along the $x$-axis from the origin of the coordinates.

Inserting the gaussian ansatz into the expression for $\mathcal{L}$ and integrating
we obtain the average Langrangian of the problem, $L=\intl \mathcal{L}\;\ud x$.
Using the Euler-Langrange equations with variational parameters $A$, $b$, $x_0$,
$\alpha$, $\beta$ and $\gamma$ we obtain a set of coupled equations describing the
evolution of these parameters, namely

\begin{subequations}
\begin{gather}
A_z=-d_0\alpha A\\
b_z=-4d_0\alpha b\\
x_{0z}=d_0(2\alpha x_0+\beta) \label{x0.eq}\\
\alpha_z=\frac{1}{2}(-\sqrt{2}n_2A^2b+4d_0b^2+n_1-4d_0\alpha^2) \label{alpha}\\
\beta_z=\sqrt{2}n_2A^2bx_0-4d_0b^2x_0-2d_0\alpha\beta\\
\gamma_z=\frac{1}{8}(5\sqrt{2}n_2A^2-8d_0b-4\sqrt{2}n_2A^2bx_0^2\nonumber\\
\hspace{3.5cm}+16d_0b^2x_0^2-4d_0\beta^2)
\end{gather}
\label{system}
\end{subequations}
By dividing the first two equations and integrating, we obtain $A^4=E_0 b$, where
$E_0$ is the energy of the system at $z=0$. This is equivalent to the conservation
of energy of the system. Indeed, Eq. (\ref{nls}) is integrable and has an infinite
number of conservation laws the first of which characterizes the system's energy,
i.e. $\intl |\psi|^2\;\ud x=E_0$. When we substitute for the profile of Eq.
(\ref{profile}) we obtain $A^4=E_0 b$, as above. In fact, all of Eqs.
(\ref{system}) can also be derived using conservation laws arguments.

\begin{figure}[!htbp]
    \centering
    \includegraphics[height=2in]{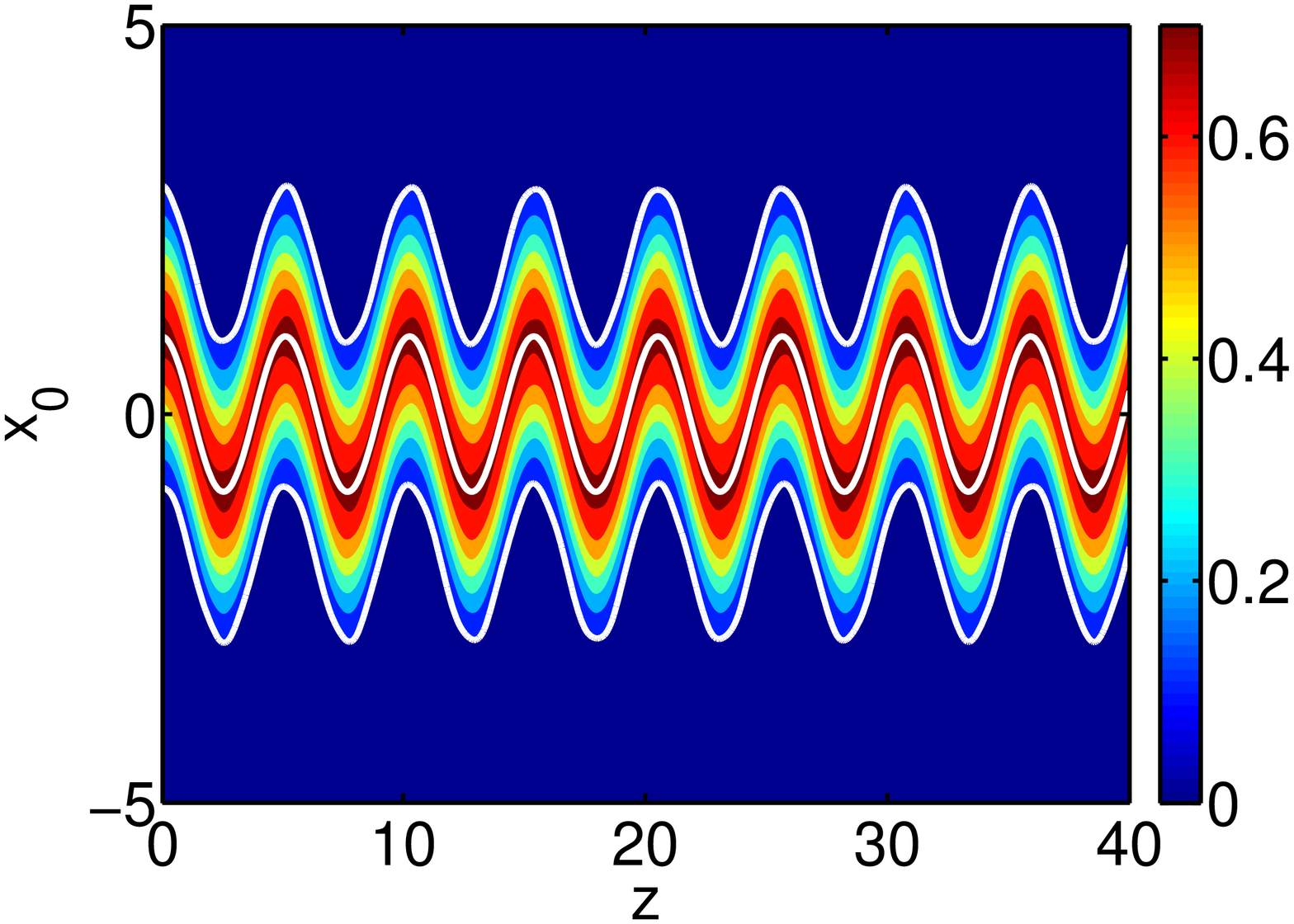}
    \includegraphics[height=2in]{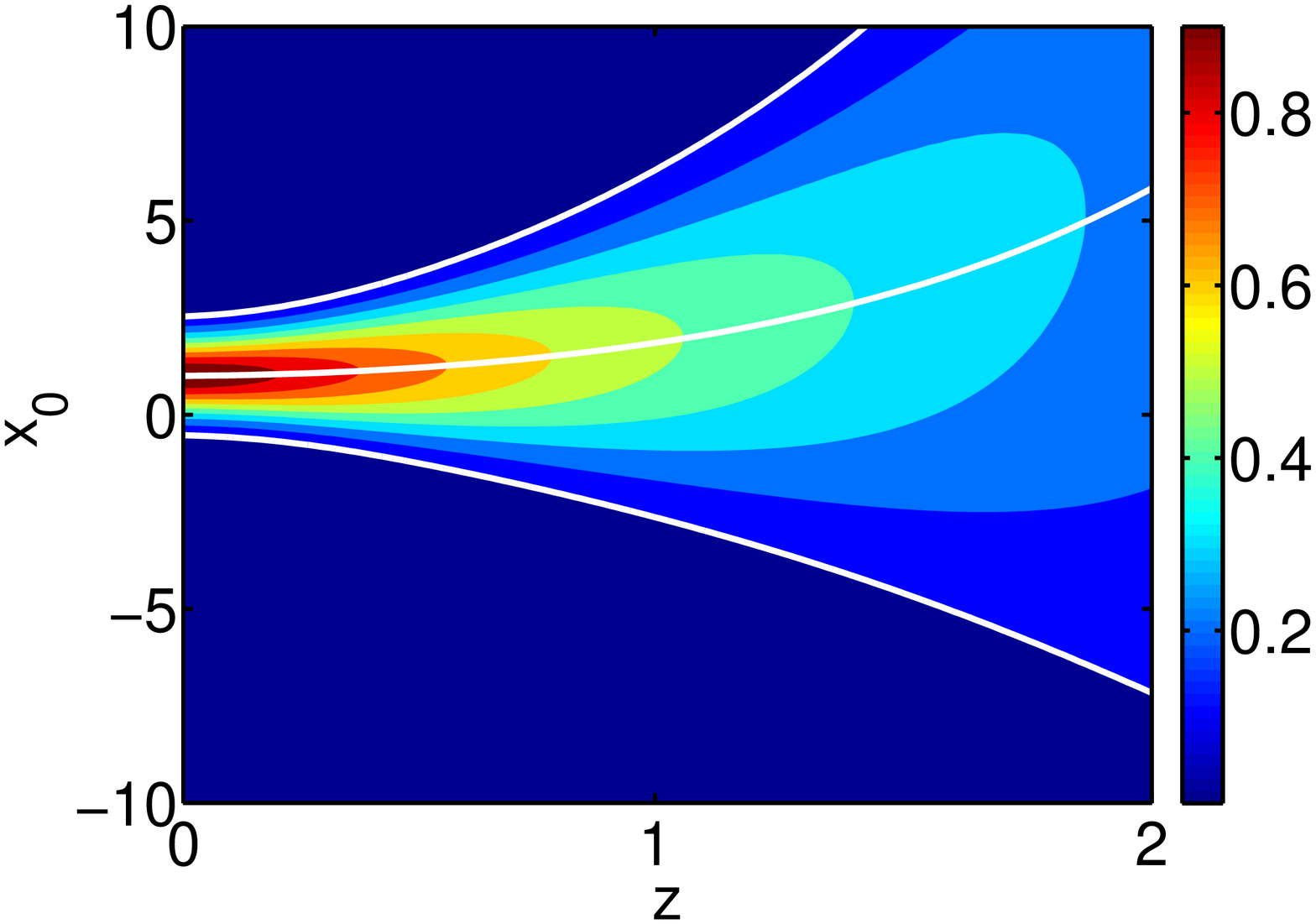}
    \caption{(Color Online) The center of a pulse oscillating with frequency $\omega=\sqrt{|d_0n_1|}$, (top)
    and exponential moving (bottom).
    The contours are the direct evolution of Eq. (\ref{nls}) and the white lines the solution of the
    uncoupled linear equation (width and center of the pulse).}
    \label{x0.fig}
\end{figure}

In general, the system of Eqs. (\ref{system}) is coupled and nonlinear. Remarkably,
however, differentiating Eq. (\ref{x0.eq}) and using the rest of the equations we
obtain the uncoupled equation
\begin{gather}
x_{0,zz}-(d_0n_1)x_0=0 \label{x0zz}
\end{gather}
The above equation demonstrates that two types of evolution can be observed. If
$d_0n_1>0$ the location of the center of the pulse moves along an exponential
trajectory. The more interesting case is the one with $d_0n_1<0$, since it suggests
an oscillatory pattern around $x_0=0$, with frequency $\omega=\sqrt{|d_0n_1|}$.
This is illustrated in, Fig. \ref{x0.fig}, where we show how the center of the
pulse evolves, with parameters $d_0=-1$, $n_1=\pm 1.5$ (top/bottom) and $x_0(0)=1$.
Hereafter, $n_2=1$ and the relative signs between diffraction and nonlinearity will
be controlled by $d_0$. Also, note that depending on the initial conditions on
$x_0(z)$ the oscillation will undergo a sine $(x_0(0)=0,\,x_{0,z}\neq 0)$ or cosine
$(x_0(0)\neq 0,\,x_{0,z}= 0)$ oscillation.
This resembles linear propagation of light in a parabolic-index optical fiber,
where light is guided if the refractive index is maximal in the center, provided
diffraction is positive. A parabolic index with the minimal value of the index in
the center would lead to defocusing of the beam. In the case where light is guided,
we can expect that different propagation modes will be allowed in the waveguiding
structure induced by the graded index. Therefore, if the input profile is not
perfectly symmetric, at least one antisymmetric mode will be excited, with a
slightly different propagation constant than the fundamental mode. The beating
between the main antisymmetric mode and the main symmetric mode will lead to a
periodic change in position of the center of the beam along the propagation.

\begin{figure}[tbp]
    \centering
    \includegraphics[height=2in]{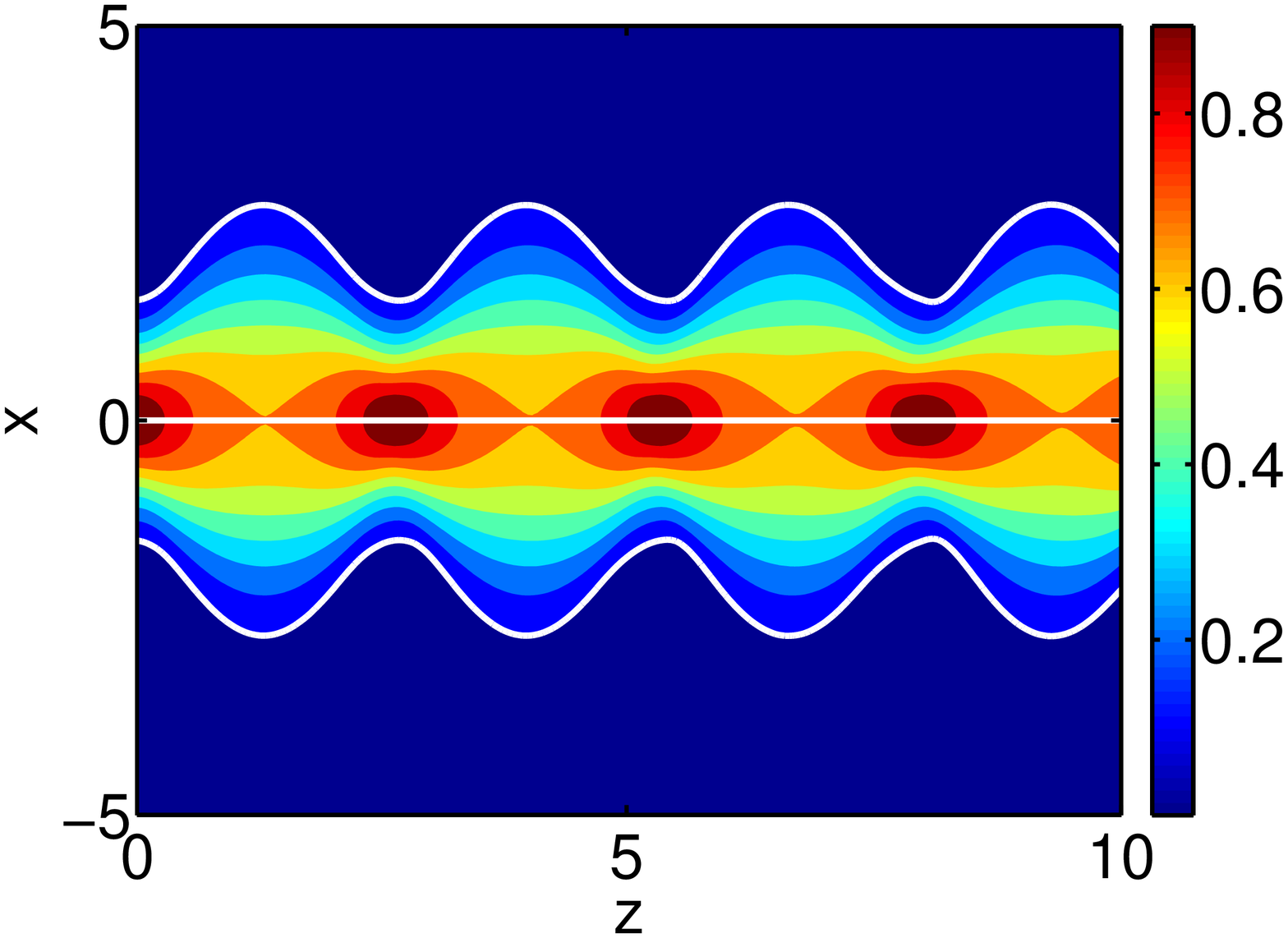}
    \includegraphics[height=2in]{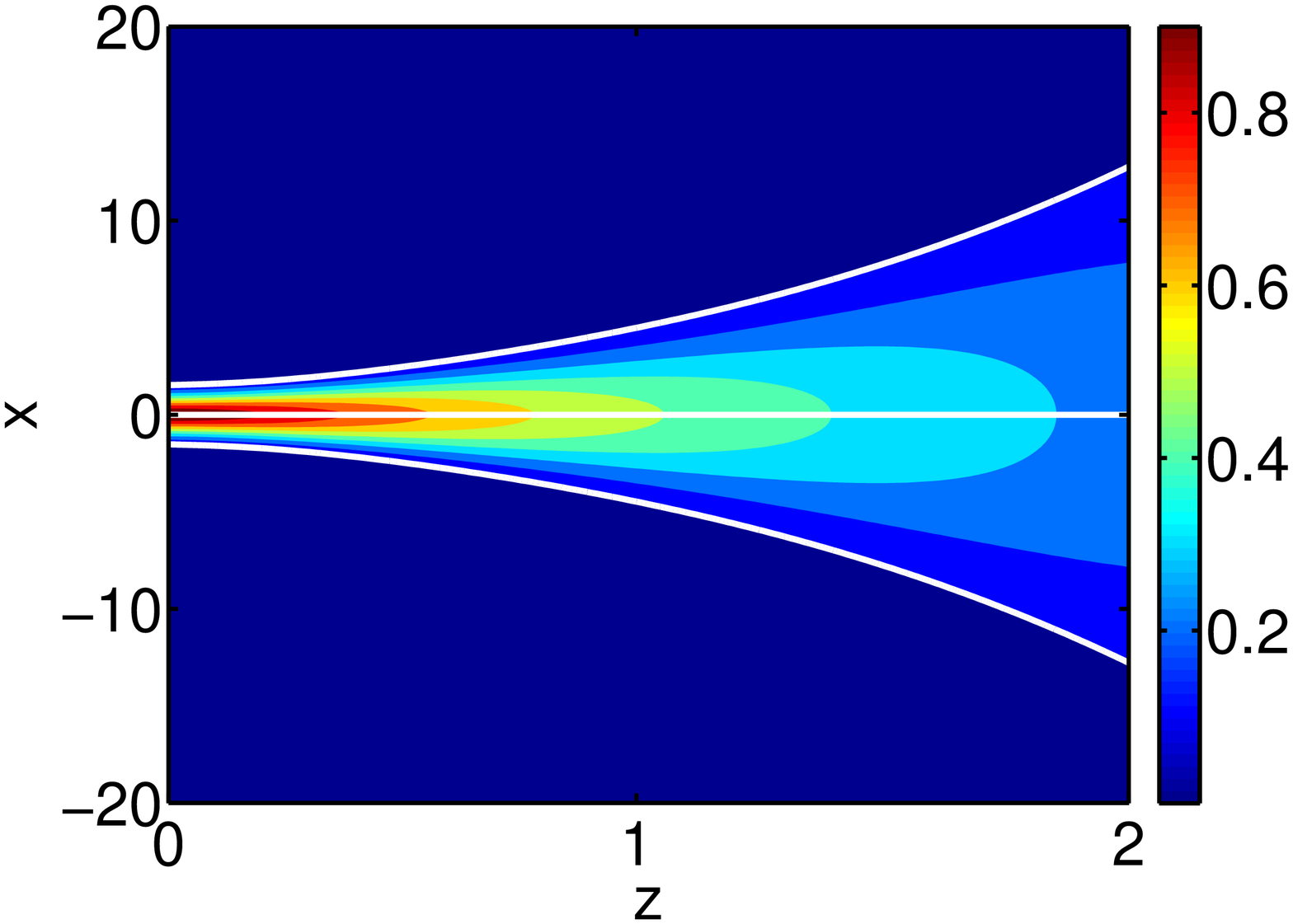}
    \caption{(Color Online) The contour plot of the complete evolution of a unit gaussian pulse with $d_0=-1$,
    $n_1=1.5$ (opposite signs) and $x_0=0$ (top). The contour plot of the complete evolution of a unit gaussian
    pulse with $d_0=-1$,
    $n_1=-1.5$ (same signs) and $x_0=0$ (bottom). The white line represents the solution of Eqs. (\ref{system}).}
    \label{A.contour}
\end{figure}

In a similar manner, a breathing behavior may be observed if one is to analyze the
propagation of a pulsed beam through the simple nonlinear device composed of a
waveguide of certain thickness and refractive index $\tilde{n}_1(\neq n_1)$,
surrounded by a linear substrate of index $n_1$ and a nonlinear cover with
Kerr-type nonlinearity of the form $n=n_1+n_2|\psi|^2$. To demonstrate, we show in
Fig. \ref{A.contour}-(top) the propagation of a unit gaussian under Eq. (\ref{nls})
with $d_0=-1$ and $n_1=1.5$. By changing the values of these parameters the
qualitative behavior of the oscillations are also changing. Moreover, if these two
parameters share the same sign no oscillations are observed, as shown in Fig.
\ref{A.contour}-(bottom). In this case the initial pulse (i.e. unit gaussian with
equation parameters $d_0=-1$ and $n_1=-1.5$) is decaying fast.

\begin{figure}[!htbp]
    \centering
    \includegraphics[height=2in]{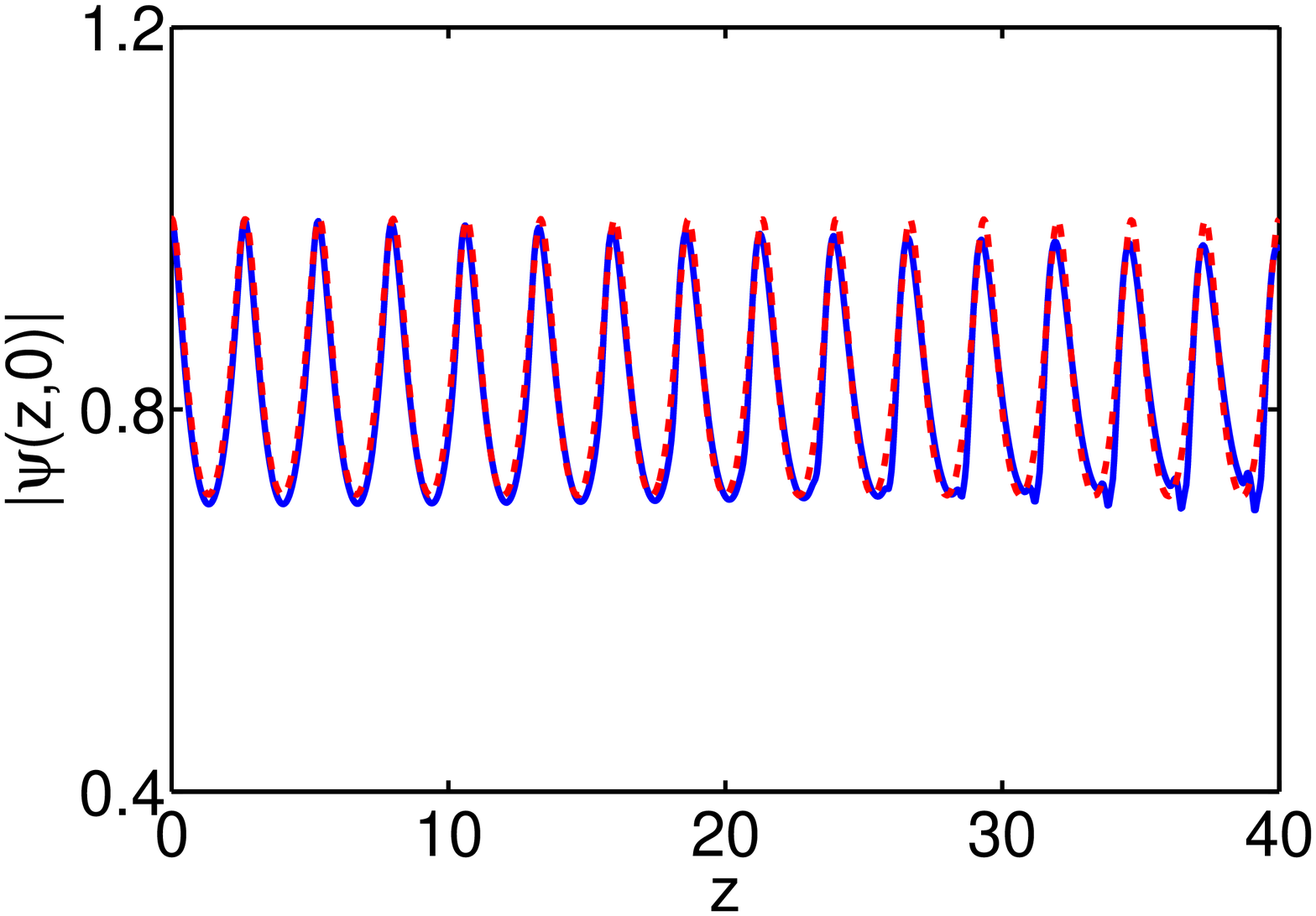}
    \includegraphics[height=2in]{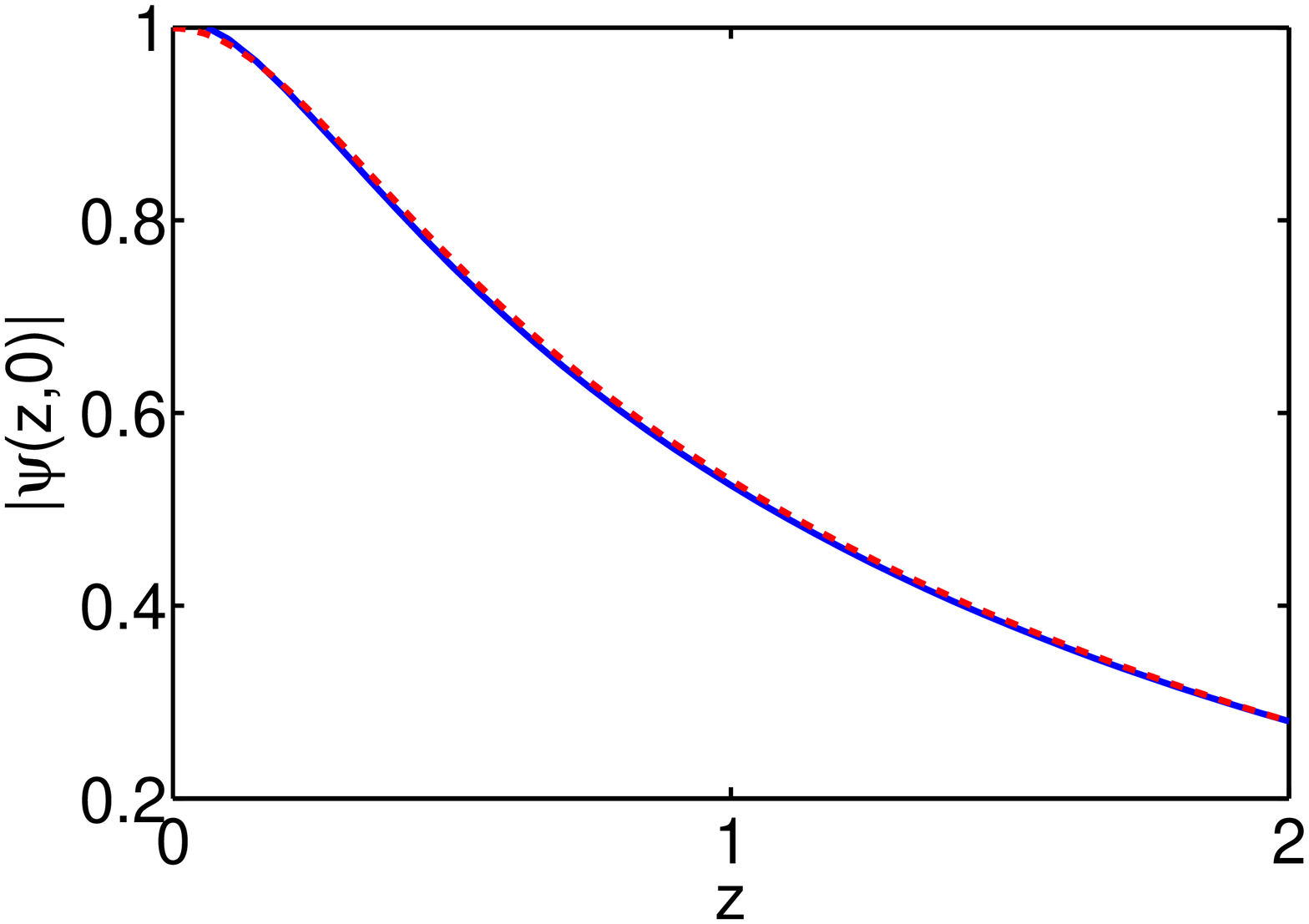}
    \caption{(Color Online) The evolution of the amplitude of a unit gaussian with
    $n_1=-1.5$. The diffraction is $d_0=\pm 1$ (top/bottom) and  and $x_0=0$.}
    \label{A.fig}
\end{figure}

\begin{figure}[!htbp]
    \centering
    \includegraphics[height=2in]{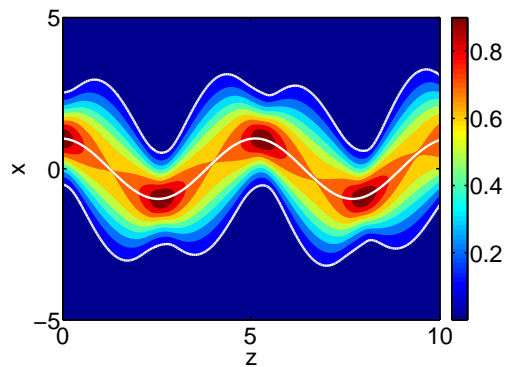}
    \caption{(Color Online) The complete evolution of a unit gaussian centered initially
    at $x_0=1$ and $n_1=1.5$, $d_0=-1$. The white line is the solution of Eqs. (\ref{system}).}
    \label{gen.fig}
\end{figure}

To further illustrate this breathing we plot in Fig. \ref{A.fig} the evolution of
the amplitude $A(z)$ of the pulse for two different values of $n_1$ starting from a
unit gaussian.
The analysis for this case is somewhat more involved than in the previous case
since the equations do not uncouple in a trivial way. Indeed, differentiating Eq.
(\ref{alpha}) and using the rest of the set we obtain a second order nonlinear,
coupled equation for $\alpha(z)$ that reads
\[
\alpha_{zz}+10d_0\alpha_z-8d_0^2\alpha^3+20d_0^2\alpha^2+4d_0^2b^2\alpha-3d_0n_1\alpha=0
\]
The oscillatory nature of the equation becomes apparent from its linear part which
can be simplified if $b^2\ll 1$ to
\[
\alpha_{zz}-(3d_0n_1)\alpha=0
\]
Again the oscillations exist if $d_0n_1<0$ and a good estimate for their frequency
is $\omega'=\sqrt{3|d_0n_1|}$. Notice that $n_2$ (nonlinearity) comes into the
system in higher-order and the difference with the frequency of the pulse's center
(also apparent in Fig. \ref{gen.fig}).

We finally briefly discuss the evolution of a unit gaussian originally dislocated
at $x_0=1$, as shown in Fig. \ref{gen.fig}.
Since the equation for $x_0$ can be uncoupled the oscillations of the center of the
pulse do not effect the breathing of the rest of the pulse's parameters. This means
that the propagation of the center of the pulse is independent of the breathing
that may occur due to changes in the refractive index of the medium.

To conclude, we demonstrated that under certain conditions, namely that diffraction
and refractive index have opposite signs, pulses propagating in parabolic index
optical waveguides exhibit oscillatory patterns. While the pattern for the
dislocation of the pulse is simple and described by a linear equation the rest of
the pulse's parameters are described by coupled, nonlinear equations. In the first
case, we provided the exact frequency of the oscillation while an estimate based on
the linear part of the equation was provided for the latter.

We wish to thank D.J. Frantzeskakis for many useful discussions while preparing
this manuscript and the anonymous reviewers for many clarifying remarks.

\bibliographystyle{osajnl}

\begin{thebibliography}{10}
\newcommand{\enquote}[1]{``#1''}

\bibitem{longhi}
S.~Longhi and D.~Janner, \enquote{Self-focusing and nonlinear periodic beams in
  parabolic index optical fibres,} J. Opt. B: Quantum Semiclass. Opt.
  \textbf{6}, S303–S308 (2004).

\bibitem{michinel}
H.~Michinel, \enquote{Pulsed nonlinear surface waves and soliton emission at
  nonlinear graded index waveguides,} Opt. Quant. Elec. \textbf{30}, 79--97
  (1998).

\bibitem{karlsson}
M.~Karlsson, D.~Anderson, and M.~Desaix, \enquote{Dynamics of self-focusing and
  self-phase modulation in a parabolic index optical fiber,} Opt. Lett.
  \textbf{17}, 22--24 (1992).

\bibitem{longhi2}
S.~Longhi, \enquote{Modulational instability and space time dynamics in
  nonlinear parabolic-index optical fibers,} Opt. Lett. \textbf{28}, 2363--2365
  (2003).

\bibitem{agrawal2}
S.~Raghavan and G.~Agrawal, \enquote{Spatiotemporal solitons in inhomogeneous
  nonlinear media,} Opt. Comm. \textbf{180}, 377–382 (2000).

\bibitem{bec2}
P.~Kevrekidis, D.~Frantzeskakis, and R.~Carretero-Gonz\'alez (eds), \emph{Emergent nonlinear
  phenomena in {Bose-Einstein} condensates: Theory and experiment} (Springer,
  2007).

\bibitem{bec3}
R.~Carretero-Gonz\'alez, D.~Frantzeskakis, and P.~Kevrekidis,
  \enquote{Nonlinear waves in {Bose-Einstein} condensates: Physical relvance
  and mathematical techniques,} Nonlinearity \textbf{21}, R139--R202 (2008).

\bibitem{abdull1}
F.~Abdullaev, R.~Galimzyanov, and K.~Ismatullaev, \enquote{Collective
  oscillations of a quasi-one-dimensional bose condensate under damping,} Phys.
  Lett. A \textbf{357}, 48--53 (2006).

\bibitem{agrawal}
S.~Ponomarenko and G.~Agrawal, \enquote{Optical similaritons in nonlinear
  waveguides,} Opt. Lett. \textbf{32}, 1659--1661 (2007).

\bibitem{wu}
L.~Wu, J.-F. Zhang, L.~Li, Q.~Tian, and K.~Porsezian, \enquote{Similaritons in
  nonlinear optical systems,} Opt. Express \textbf{16}, 6352--6360 (2008).

\end{thebibliography}

%
\end{document}